# Conditions for the spin-spiral state in itinerant magnets

M.A. Timirgazin [a], A.K. Arzhnikov [b]

Physical-Technical Institute,

Ural Branch of Russian Academy of Sciences,

Kirov str. 132, Izhevsk 426001, Russia

[a]timirgazin@gmail.com, [b]arzhnikov@otf.pti.udm.ru

**Abstract.** The spin-spiral (SS) type of magnetization is studied with the Hubbard model. Consideration of noncollinearity of the magnetic moments results in a phase diagram which consists of regions of the SS and paramagnetic states depending on the number of electrons $n_{el}$ and the parameter $U/t$ ($U$ is the Hubbard repulsion, and $t$ is an overlap integral). A possibility of stabilization of the SS state with three nonzero components of magnetic moment is considered.

**Introduction**

Conditions of realization of various noncollinear magnetic states (NCMS) in itinerant metals are not well studied at present. The spin-spiral (SS) state is a special case of NCMS in which the magnetic order presents a spiral wave. The SS type of magnetization was actively theoretically investigated at the beginning of the 90s in the context of high-$T_C$ superconductors. The studies were performed mostly in the framework of the Hubbard model, and their general result was that the SS state is realized in the system in a wide range of parameters of the model [1-5]. However, quantitative results of these works do not agree with each other in many ways (see Ref. [3] and Ref. [4], for example).

In the last years new interesting experimental and theoretical data concerning the SS type of magnetization were revealed. It was found that the SS state is most likely realized in Fe-Al alloys for certain concentrations [6-8]. Authors of Ref. [9] state that even typical ferromagnetic materials (Fe, Co, Ni) should be possible to stabilize in the SS order. Recently the data appear which indicate appearance of the giant magnetic resistance effect in the systems with the SS state [10]. Thus a requirement has arisen for more careful investigation of conditions for the SS state in itinerant magnets.

We perform calculations of the SS state phase diagram for a plane square lattice taking into account all possible orientations of the SS wave vector. Moreover we study the possibility of the SS state stabilization with magnetic moments which lay off the $xy$-plane, i.e. they have a nonzero $z$-component as well.

**Model and results**

The Hamiltonian in the Hubbard model is:

$$H = \frac{1}{N} \sum_{\mathbf{j},\mathbf{д},\sigma} t c^+_{\mathbf{j},\sigma} c_{\mathbf{j}+\mathbf{д},\sigma} + \frac{U}{2N} \sum_{\mathbf{j},\sigma} c^+_{\mathbf{j},\sigma} c^+_{\mathbf{j},-\sigma} c_{\mathbf{j},-\sigma} c_{\mathbf{j},\sigma}. \tag{1}$$

Here and hereinafter we follow the designations from [11]. $c^+_{\mathbf{j},y}$ and $c_{\mathbf{j},y}$ are the creation and destruction operators of electron at site $\mathbf{j}$ with the spin direction y, д is a vector which connects the site $\mathbf{j}$ with its neighbors, $N$ is the number of sites in the lattice. Assuming that the directions of quantization axes $\mathbf{e_j}$ can be different depending on site, the Hamiltonian can be rewritten in the form:

$$H = \frac{1}{N}\sum_{\mathbf{j},\mathbf{д},\sigma} t c^+_{\mathbf{j},\sigma} c_{\mathbf{j}+\mathbf{д},\sigma} + \frac{1}{4}Un_{el}^2 - \frac{U}{N}\sum_{\mathbf{j}}(\mathbf{e_j S_j})^2, \quad (2)$$

where $\mathbf{S_j}$ is a spin at site $\mathbf{j}$, $n_{el}$ is a number of electrons per one site. Solving the task with the Hamiltonian (2) is a difficult problem, and only few mathematically exact results have been studied for generality [11]. At the same time, very helpful are the results obtained in the mean-field approximation which is also used here. After some identical transformations we rewrite the Hamiltonian in the form:

$$H = H_0 + H_{int} \quad (3)$$

$$H_0 = U\left(M^2 + \frac{n_{el}^2}{4}\right) + \frac{1}{N}\sum_{\mathbf{j},\mathbf{д},\sigma} t c^+_{\mathbf{j},\sigma} c_{\mathbf{j}+\mathbf{д},\sigma} - 2\frac{UM}{N}\sum_{\mathbf{j}}(\mathbf{e_j S_j})$$

$$H_{int} = -\frac{U}{N}\sum_{\mathbf{j}}(\mathbf{e_j S_j} - M)^2,$$

where M is a magnitude of magnetic moment at site.

Direct calculations show that $_0\langle H_{int}\rangle_0 = 0$, where $|\rangle_0$ is the ground state of the Hamiltonian $H_0$. Neglecting $H_{int}$, we make a mean-field approximation.

We assume that the magnetization distribution over sites looks like a spin spiral with magnetic moments which rotate in $xy$-plane and incline on angle $\psi$ with respect to $z$-axis, i.e. the quantization axes has the following form:

$$\mathbf{e_j} = \mathbf{e}_x \sin\psi \cos\mathbf{QR_j} + \mathbf{e}_y \sin\psi \sin\mathbf{QR_j} + \mathbf{e}_z \cos\psi.$$

Making a transition to the Fourier representation we rewrite the Hamiltonian $H_0$ in a more convenient form:

$$H_0 = U\left(M^2 + \frac{n_{el}^2}{4}\right) + \frac{1}{N}\sum_{\mathbf{k}}\varepsilon_{\mathbf{k}}^{0\uparrow} c^+_{\mathbf{k},\uparrow} c_{\mathbf{k},\uparrow} + \frac{1}{N}\sum_{\mathbf{k}}\varepsilon_{\mathbf{k+Q}}^{0\downarrow} c^+_{\mathbf{k+Q},\downarrow} c_{\mathbf{k+Q},\downarrow} - \frac{1}{N}UM\sin\psi\sum_{\mathbf{k}}\{c^+_{\mathbf{k},\uparrow} c_{\mathbf{k+Q},\downarrow} + c^+_{\mathbf{k+Q},\downarrow} c_{\mathbf{k},\uparrow}\},$$

(4)

where:

$$\varepsilon_{\mathbf{k}}^{0,\uparrow} = \varepsilon_{\mathbf{k}}^0 - UM\cos\psi, \qquad \varepsilon_{\mathbf{k+Q}}^{0,\downarrow} = \varepsilon_{\mathbf{k+Q}}^0 + UM\cos\psi, \qquad \varepsilon_{\mathbf{k}}^0 = t\sum_{\mathbf{д}}\exp^{i\mathbf{k}\mathbf{д}},$$

$$c_{\mathbf{k},\sigma} = \frac{1}{\sqrt{N}}\sum_{\mathbf{j}} c_{\mathbf{j},\sigma}\exp^{-i\mathbf{k}\mathbf{R_j}}, \qquad c^+_{\mathbf{k},\sigma} = \frac{1}{\sqrt{N}}\sum_{\mathbf{j}} c^+_{\mathbf{j},\sigma}\exp^{i\mathbf{k}\mathbf{R_j}},$$

$$S_{\mathbf{k}}^+ = \frac{1}{N}\sum_{\mathbf{j}} S_{\mathbf{j}}^+ \exp^{-i\mathbf{k}\mathbf{R_j}} = \frac{1}{N}\sum_{\mathbf{k'}} c^+_{\mathbf{k'},\uparrow} c_{\mathbf{k'+k},\downarrow}, \qquad S_{\mathbf{k}}^- = \frac{1}{N}\sum_{\mathbf{j}} S_{\mathbf{j}}^- \exp^{i\mathbf{k}\mathbf{R_j}} = \frac{1}{N}\sum_{\mathbf{k'}} c^+_{\mathbf{k'+k},\downarrow} c_{\mathbf{k'},\uparrow}.$$

The Hamiltonian $H_0$ can be diagonalized in the operator representation

$$A_{\mathbf{k}} = c_{\mathbf{k},\uparrow}\cos\theta_{\mathbf{k}} + c_{\mathbf{k+Q},\downarrow}\sin\theta_{\mathbf{k}},$$

$$B_{\mathbf{k+Q}} = c_{\mathbf{k},\uparrow}\sin\theta_{\mathbf{k}} - c_{\mathbf{k+Q},\downarrow}\cos\theta_{\mathbf{k}},$$

$$\tan(2\theta_{\mathbf{k}}) = \frac{-UM\sin\psi}{\dfrac{\varepsilon_{\mathbf{k}}^0 - \varepsilon_{\mathbf{k+Q}}^0}{2} - UM\cos\psi}$$

(analogous expressions relate the creation operators $A^+_{\mathbf{k}}$ and $B^+_{\mathbf{k+Q}}$ with $c^+_{\mathbf{k},\uparrow}$ and $c^+_{\mathbf{k+Q},\downarrow}$):

$$H_0 = U(M^2 + \frac{n_{el}^2}{4}) + \frac{1}{N}\sum_{\mathbf{k}}\{\varepsilon_{\mathbf{k}}^A A^+_{\mathbf{k}} A_{\mathbf{k}} + \varepsilon_{\mathbf{k+Q}}^B B^+_{\mathbf{k+Q}} B_{\mathbf{k+Q}}\} \quad (5)$$

where

$$\varepsilon_{\mathbf{k}}^A = \varepsilon_{\mathbf{k}}^+ + \mathrm{sign}(\varepsilon_{\mathbf{k}}^- - UM\cos\psi)\sqrt{(\varepsilon_{\mathbf{k}}^- - UM\cos\psi)^2 + (MU\sin\psi)^2},$$

$$\varepsilon^B_{k+Q} = \varepsilon^+_k - \text{sign}(\varepsilon^-_k - UM\cos\psi)\sqrt{(\varepsilon^-_k - UM\cos\psi)^2 + (MU\sin\psi)^2},$$

$$\varepsilon^+_k = \frac{\varepsilon^0_k + \varepsilon^0_{k+Q}}{2}, \qquad \varepsilon^-_k = \frac{\varepsilon^0_k - \varepsilon^0_{k+Q}}{2}.$$

To determine magnetic order and calculate its characteristics it is necessary to solve self-consistently, at fixed parameters $U/t$, $Q$ and $\psi$, two equations relating the magnetic moment, the number of electrons and the Fermi level:

$$n_{el} = \frac{1}{N} \sum_k^{\varepsilon_F} (n^A_k + n^B_{k+Q}) \tag{6}$$

$$_0\langle S^\pm_Q\rangle_0 = M = \frac{1}{2N} \sum_k^{\varepsilon_F} (n^A_k - n^B_{k+Q})(\sin 2\theta_k \sin\psi + \cos 2\theta_k \cos\psi) \tag{7}$$

where $n^B_k = {}_0\langle B^+_k B_k\rangle_0$, $n^A_k = {}_0\langle A^+_k A_k\rangle_0$. Then a minimum of the total energy

$$E_{SS}(Q,\psi) = {}_0\langle H\rangle_0 = U\left(M^2 + \frac{n^2_{el}}{4}\right) + \frac{1}{N} \sum_k^{\varepsilon_F} (\varepsilon^A_k n^A_k + \varepsilon^B_{k+Q} n^B_{k+Q}) \tag{8}$$

over the $Q$ wave vector and the $\psi$ angle should be found.

The calculations themselves have been performed for a plane square lattice. The coordinate axes x and y are directed along the square sides, the dispersion law being determined as $\varepsilon^0_k = t(\cos(k_x) + \cos(k_y))$. Here and after, the lattice parameter, the Lande's factor and the Bohr's magneton are taken as unity. It should be mentioned that in our model the result does not depend on the direction of the spins with respect to the crystallographic axes and is determined by the $Q$ vector and the $\psi$ angle only. A search of all magnetic solutions has been conducted over the whole range of model parameters. A comparison of total energies of different solutions shows that the phase diagram consists of the paramagnetic and SS regions (Fig. 1). The calculations show that the SS phases with [Q, Q], [Q, π] and [0, Q] directions of the wave are realized. Also ferromagnetic (F), antiferromagnetic (AF) and [0, π] phases are depicted on the phase diagram though they can be considered as special cases of the phases enumerated above. Note that the AF state takes place at half filling ($n_{el} = 1$) only. No parameters were found for which the SS state with nonzero z-component of the moment is minimal by the energy ($E_{SS}(\psi)$ dependencies were studied for the points: a) $U/t = 10$, $n_{el} = 0.3$; b) $U/t = 9$, $n_{el} = 0.5$; c) $U/t = 6$, $n_{el} = 0.7$; d) $U/t = 10$, $n_{el} = 0.8$).

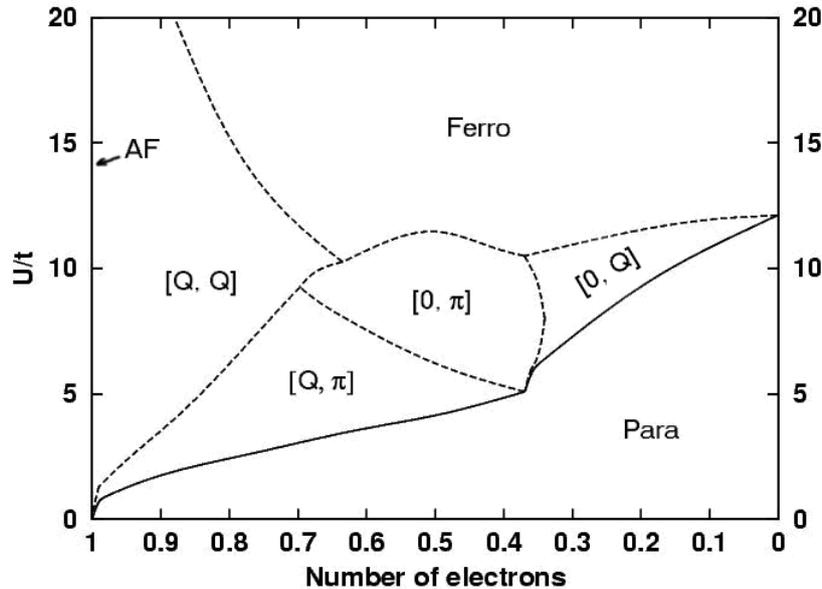

Fig.1. The phase diagram of the SS state for a plane square lattice. (The diagram is symmetric with respect to the line $n_{el} = 1$, so only half of it is shown.)

The line separating the paramagnet (P) state from the states with SS magnetic ordering is determined by the equation:

$$1/U \underset{M \to 0}{=} -\frac{1}{N} \sum_{\mathbf{k}}^{\varepsilon_F} \frac{n_{\mathbf{k}}^A - n_{\mathbf{k+Q}}^B}{\varepsilon_{\mathbf{k}}^0 - \varepsilon_{\mathbf{k+Q}}^0}, \quad (9)$$

which can be referred to as a generalized Stoner criterion for SS. The right-hand side of Eq. (9) is the SS polarization function and identical to the polarization function of the state with a spin-density wave [11].

Phase diagrams calculated previously, for example in Ref. [3], are considerably less accurate and do not reveal the [0, Q] region. Results of Ref. [4] where a stabilization of the [$Q_1$, $Q_2$] state is predicted are not correct because of inadequate calculation of the total energy in the work (a term with $M^2$ is missed).

**Summary**

1)  Our model deals with the SS state for which a magnetic moment rotates in *xy*-plane and have a nonzero *z*-component as well. We have found that any deviation of the *z*-component from zero results in increase of the total energy of the system for all the parameters studied.

2)  The phase diagram for the SS state in itinerant metals has been constructed. Our calculations were more accurate in comparison with the analogous calculations performed previously by the other authors. In contrast to the other works we have found the SS state with the [0, Q] direction of the wave to be realized in the system for certain parameters.